\documentclass[prl,twocolumn,showpacs,preprintnumbers,amsmath,amssymb,superscriptaddress]{revtex4}

\usepackage{graphicx}
\usepackage{dcolumn}
\usepackage{bm}

\begin{document}


\title{Polar Antiferromagnets Produced with Orbital-Order}

\author{Naoki~Ogawa}
 \email[]{ogawa@myn.rcast.u-tokyo.ac.jp}
\affiliation{Research Center for Advanced Science and Technology (RCAST), 
The University of Tokyo, 4-6-1 Komaba, Meguro-ku, Tokyo 153-8904, Japan}
\affiliation{CREST, Japan Science and Technology Agency,
4-1-8 Honcho, Kawaguchi, Saitama 332-0012, Japan}

\author{Yasushi~Ogimoto}
\affiliation{Research Center for Advanced Science and Technology (RCAST), 
The University of Tokyo, 4-6-1 Komaba, Meguro-ku, Tokyo 153-8904, Japan}
\affiliation{CREST, Japan Science and Technology Agency,
4-1-8 Honcho, Kawaguchi, Saitama 332-0012, Japan}
\affiliation{Fuji Electric Co., Ltd., Hino, Tokyo 191-8502, Japan}

\author{Yoshiaki~Ida}
\affiliation{Department of Applied Physics, Graduate School of Engineering, The University of Tokyo, 
7-3-1 Hongo, Bunkyo-ku, Tokyo 113-8656, Japan}

\author{Yusuke~Nomura}
\affiliation{Department of Applied Physics, Graduate School of Engineering, The University of Tokyo, 
7-3-1 Hongo, Bunkyo-ku, Tokyo 113-8656, Japan}

\author{Ryotaro~Arita}
\affiliation{Department of Applied Physics, Graduate School of Engineering, The University of Tokyo, 
7-3-1 Hongo, Bunkyo-ku, Tokyo 113-8656, Japan}
\affiliation{PREST, Japan Science and Technology Agency,
4-1-8 Honcho, Kawaguchi, Saitama 332-0012, Japan}

\author{Kenjiro~Miyano}
\affiliation{Research Center for Advanced Science and Technology (RCAST), 
The University of Tokyo, 4-6-1 Komaba, Meguro-ku, Tokyo 153-8904, Japan}
\affiliation{CREST, Japan Science and Technology Agency,
4-1-8 Honcho, Kawaguchi, Saitama 332-0012, Japan}

\date{\today}

\begin{abstract}
Polar magnetic states are realized in pseudocubic manganite thin films fabricated on high-index substrates, in which a Jahn-Teller (JT) distortion remains an active variable. Several types of orbital-orders were found to develop large optical second harmonic generation, signaling broken-inversion-symmetry distinct from their bulk forms and films on (100) substrates. The observed symmetry-lifting and first-principles calculation both indicate that the modified JT $q_{2}$ mode drives Mn-site off-centering upon orbital order, leading to the possible cooperation of ``Mn-site polarization'' and magnetism.
\end{abstract}

\pacs{77.55.Nv, 77.80.bn, 78.47.jh}
\maketitle
Cross-correlated controls of electronic functionalities, such as magnetic responses induced by an electric field and {\it vice versa}~\cite{TKimura}, are of particular importance in oxide electronics. Although there exist many bulk compounds showing collective spin, charge, and orbital orders (SO, CO, and OO)~\cite{YTokura1}, most of them lack either electric polarization itself or sufficient couplings between electric fields and respective degrees of freedom~\cite{SWCheong}. Therefore incorporation of engineered polarization in magnetic perovskites, with effectual cross-couplings, should renovate the nature of functional oxides. To modify the electronic symmetry of these orders, application of epitaxial strain~\cite{YKonishi} and the formation of interfaces~\cite{JHe} are promising.

In pseudocubic perovskites ($AB$O$_{3}$), proper ferroelectricity is usually induced by the atomic displacement of lone-pair active $A$-sites (e.g., BiMnO$_{3}$) and/or $B$-sites with $d^{0}$-ness (e.g., BaTiO$_{3}$). However, empirical incompatibility of ferroelectricity with magnetism at the $B$-sites~\cite{NAHill} practically directs the researchers to seek for the polarization without $B$-site displacements; geometrical interactions of spins and orbitals~\cite{TKimura,CJia}, CO with asymmetric charge distribution~\cite{DVEfremov,YTokunaga,NIkeda}, octahedral rotations~\cite{VGopalan,EBousquet}, and hybrid~\cite{NABenedek}.

The proper ferroelectricity in strained cubic manganites, which directly couples to $B$-site spins, has been predicted from first principles calculations~\cite{SBhattacharjee,JMRondinelli,EBousquet2}. Whereas the first-order JT distortion preserves centrosymmetry of the $B$O$_{6}$ octahedra, the second-order terms can drive the $B$-site off-centering, which is favored in the expanded lattice in relation to octahedral rotations. Following this scenario, high-pressure synthesized Sr$_{1-x}$Ba$_{x}$MnO$_{3}$(SBMO) has been reported recently to show ferroelectricity~\cite{HSakai}. However, these predictions are quite limited in both material selection and lattice modification, where tensile (epitaxial) strain has been considered only in a tetragonal way, excluding the possibility of additional symmetry-engineering.  

Here we introduce {\it shear-mode} stress as a general route to induce $B$-site displacements in pseudocubic perovskites. By using (110)-oriented substrates and doped manganites with degenerate $e_{g}$ orbitals, we can activate JT distortions at the OO transition in thin films, which induce enough shear-mode stress in the octahedra favoring the polar lattice motion. Three examples of prototypical manganites with distinct ground states are presented; Pr$_{0.5}$Sr$_{0.5}$MnO$_{3}$(PSMO), Pr$_{0.5}$Ca$_{0.5}$MnO$_{3}$(PCMO), and Nd$_{0.5}$Sr$_{0.5}$MnO$_{3}$(NSMO); with spin-A-type antiferromagnetic (AFM) state and $d_{x^{2}-y^{2}}$ OO for the former, and with spin-CE-type AFM, checker-board type and $d_{3x^{2}-r^{2}/3y^{2}-r^{2}}$ charge-orbital order (COO) for the latter two~\cite{YWakabayashi}. The manganite films are fabricated on the (110) faces of (LaAlO$_{3}$)$_{0.3}$(SrAl$_{0.5}$Ta$_{0.5}$O$_{3}$)$_{0.7}$(LSAT) and SrTiO$_{3}$(STO) substrates~\cite{YOgimoto}. Indeed we revealed the emergence of polar magnetic phases in these films, and characterized their symmetry in detail by using optical second harmonic generation (SHG).

Electronic properties of coherently-grown films can be different from those of bulk single crystals~\cite{YKonishi}. For example, first-order phase transitions in manganites are usually suppressed by the biaxial lattice clamping to the (001)-oriented substrates. In the present study, the OO transitions are made possible through the modified JT $q_{2}$ mode constrained on (110) substrates~\cite{YOgimoto}, where the two axes of the film will change in a monoclinic (or parallelogram) way keeping the cell volume~\cite{YWakabayashiL}. Note that single crystalline films with distinct phase transitions are paramount to understand the critical size of respective degrees of freedom, which is essential to realize functional nano-structures.

\begin{figure}
\includegraphics[scale=0.62]{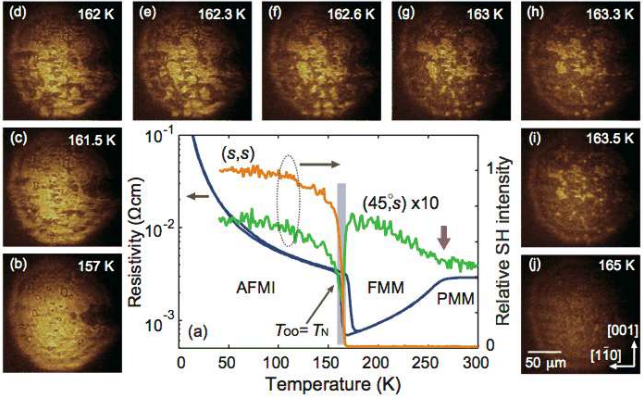}
\caption{\label{fig1} (a) Transport and SHG for the 25-nm-thick PSMO/LSAT(110) film. Successive phase transitions can be seen; from paramagnetic metal (PMM), to ferromagnetic metal (FMM), and to spin-A-type AFM insulator (AFMI) around 160 K upon cooling. The resistivity remains relatively small at lower temperature due to the spin-A-type [AFM-coupled FMM planes] nature. A vertical arrow indicates the PMM to FMM transition ($\sim$260 K). (b-j), Evolution of the orbital-ordered (OO) domains around $T_{\rm OO}$=$T_{\rm N}$ [indicated with shadow in (a)] observed with ($s,s$) polarization. Bright regions are polar (SH active), which are concomitantly in the $d_{x^{2}-y^{2}}$ OO and AFM phases.}
\end{figure}

\begin{figure}
\includegraphics[scale=0.6]{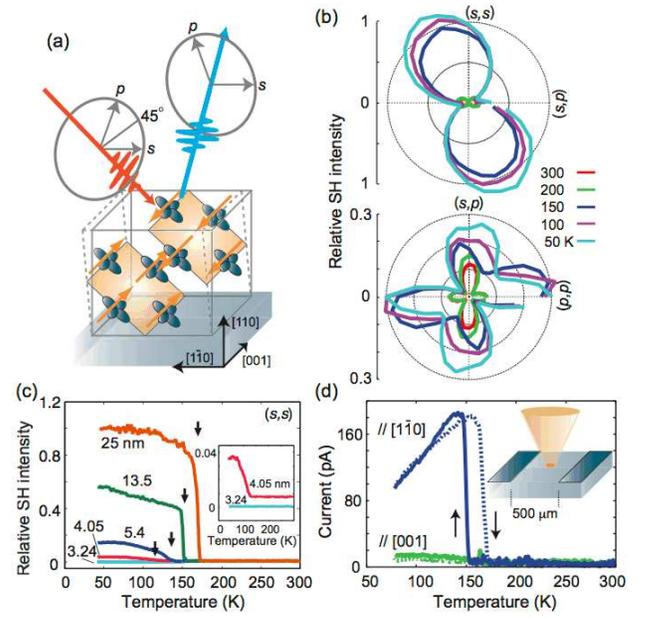}
\caption{\label{fig2} (a) A schematic of SH polarimetry and spin-A-type $d_{x^{2}-y^{2}}$ OO in the PSMO/LSAT(110) film with the experimental coordinate. The orbitals lie in the (100) or (010) plane, and orange arrows indicate the expected spin structures. The shear-mode (monoclinic) JT distortion at the transition is indicated with dotted lines. (b) SH polarization patterns of a 25-nm-thick film for the $s$-polarized ($\parallel$[1$\bar{1}$0]) fundamental photons (top) and for the $p$-polarized SH photons (bottom). The incident plane is parallel to (1$\bar{1}$0), as shown in (a). (c) ($s,s$) polarized SHG from PSMO/LSAT(110) films with different thicknesses. (d) Pyroelectric current measured on a 13.5-nm-thick film with laser intensity modulation ($\lambda =$ 660 nm, sinusoidally modulated at 1 kHz, 1.8 mW on average) focused on a 35 $\mu$m $\phi$ spot between the electrodes gap (500 $\mu$m). Inset illustrates the experimental geometry.}
\end{figure}

The characterization of symmetry and detection of small atomic displacements in oxide thin films are still challenging, even with x-rays from synchrotron radiation. Polarized SHG has many advantages including the sensitivity to the electronic symmetry at the thickness of a single unit cell. It is also possible to image polar, AFM, and toroidal domains~\cite{BBVAken} down to the diffraction limit.
Figure~\ref{fig1}(a) shows the electronic phases of a PSMO film on an LSAT(110) substrate. The $\rho$-$T$ characteristics are quite similar to that of a bulk single crystal~\cite{YTomioka}, and the metal-to-insulator (MI) transitions can be further enhanced by reducing the film thickness (not shown). According to the study of superlattice reflections~\cite{YWakabayashi}, the MI transition was confirmed to be into a $d_{x^{2}-y^{2}}$ OO phase without any CO. Pseudocubic manganites with tetragonal or orthorhombic structures keep a center of inversion in their bulk crystals, thus are nonpolar and SH inactive~\cite{MFiebig}. In contrast, we found strong SHG in the OO phase [Fig.~\ref{fig1}(a)] with ($s,s$) polarization, where the first letter in the parentheses refers to the polarization of the incident fundamental photons and the second letter to that of the second-harmonic, respectively [see Fig.~\ref{fig2}(a) for the coordinates. The details of the SHG measurement can be found elsewhere~\cite{NOgawa}]. The OO transition is also discernible in (45$^{\circ},s$) polarization as a dip, which is a result of interference with the surface SHG.

The observed ($s,s$) signal is strong enough and background free to perform SHG imaging. Figures~\ref{fig1}(b-j) display domain structures emerged across the phase transition. The OO and concomitant AFM regions appear bright in the SH microscopy, ensuring that the SH is not from domain walls or other defects. Contrary to the case of spin-CE-type manganites in which nanoscale phase separation is present~\cite{KLai}, patch-like domains with the size of 1$\sim$10 $\rm \mu$m are resolved, which appear and disappear not gradually but instantly under cooling/warming processes. 

Typical SH polarization patterns of the PSMO film are shown in Fig.~\ref{fig2}(b). Above the OO transition temperature ($T_{\rm OO}=T_{\rm N}\sim$160 K, $T_{\rm N}$; N\'{e}el temperature), the patterns indicate the electronic symmetry of $mm2$ being consistent with that of the (110) surface. At $T_{\rm OO}=T_{\rm N}$, large $s$-polarized SHG (along [1$\bar{1}$0]) emerges, signaling a symmetry-breaking, which interferes with the surface $mm$2 signal and rotates the SH polarization [Fig.~\ref{fig2}(b) top panel]. The $p$-polarized SHG is also modified due to the same symmetry-lifting (bottom panel). By analyzing the SH polarization measured with several optical geometries, it is confirmed that this $p$-polarized SHG is from the [110]-oriented polarization and there is no [001]-oriented components. By assuming the monoclinic distortion of the film~\cite{YWakabayashiL}, we conclude that the symmetry in the OO phase is consistent with point group ${m}$~\cite{RRBIRSS} with the monoclinic axis along [001] [see Fig.~\ref{fig2}(a)]. The surface SHG ($mm2$) was found to be nearly temperature independent. Figure~\ref{fig2}(c) compares the ($s,s$) polarized SHG for several PSMO/LSAT(110) films with different thicknesses. The transition temperatures decrease consistently with those determined by the transports (not shown), and the signal intensity at low temperature indicates a volumetric behavior under the influence of coherence length of reflection SHG.

To confirm its polar character, pyroelectric current was measured along two in-plane axes by intensity modulation heating with a diode laser [Fig.~\ref{fig2}(d)]. Surprisingly large current of the order of 100 pA was detected only along [1$\bar{1}$0], although the transport and magnetization are nearly isotropic in these films~\cite{YOgimoto}.  

\begin{figure}
\includegraphics[scale=0.57]{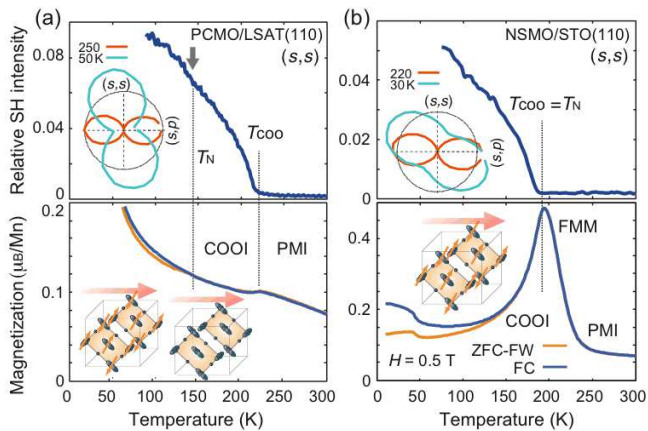}
\caption{\label{fig3} SHG and magnetic properties of (a) PCMO/LSAT(110) ($t=$ 26 nm) and (b) NSMO/STO(110) ($t=$ 40 nm) thin films. (Upper panels) Emergence of SHG with ($s,s$) polarization at $T_{\rm COO}$. Insets show changes in the SH polarization patterns for the $s$-polarized ($\parallel$[1$\bar{1}$0]) fundamental photons across the transition. (Lower panels) Magnetization-temperature curves indicating respective electronic phases. PCMO: paramagnetic insulator (PMI), charge/orbital-ordered insulator (COOI), and CE-type AFM spin-order (SO) (upon cooling). NSMO: PMI, ferromagnetic metal (FMM), and COOI with CE-type SO. Insets illustrate the expected COO/SO structures with the direction of observed SH polarization. A short arrow in the upper panel of (a) indicates a small kink in the SHG reflecting the SO transition to AFM.}
\end{figure}

In addition to the PSMO films, PCMO and NSMO films on (110) substrates are characterized to prove the universality of the modified JT deformation to induce polarization. These compounds are carefully chosen to clarify the symmetry-breaking mechanism and to compare different OOs ($d_{x^{2}-y^{2}}$ to $d_{3x^{2}-r^{2}/3y^{2}-r^{2}}$), SOs (A-type to CE-type), and substrate materials. Figure~\ref{fig3} illustrates the respective electronic phases and SHG. Although both films have the same spin-CE-type ground states [Fig.~\ref{fig3} insets], the former shows the COO and the SO at different temperatures, thus was used to single out the role of SOs. For both films we detected ($s,s$) polarized SHG below $T_{\rm COO}$; again the symmetry is lifted from their bulk forms~\cite{MFiebig}. The microscopy imaging here was not feasible, because the SH signals are more than 10 times smaller than that of the PSMO film, and domains for the spin-CE-type order is presumably small~\cite{KLai}. Moreover, the temperature dependences are rather gradual, which clearly differ from the lattice modulation at OO~\cite{YWakabayashiL}, revealing their electronic contributions. These characteristics point to the difference between the $d_{x^{2}-y^{2}}$ and $d_{3x^{2}-r^{2}/3y^{2}-r^{2}}$ OOs, in which the former develops long-range order and shows homogeneous transitions~\cite{YUozu}. Nevertheless, the changes in their electronic symmetry are similar to that of PSMO, surprisingly, as can be seen in the rotation of SH polarization (Fig.~\ref{fig3} insets). Here the rotation angle is a complex function of the refractive index and the phase of the second-order susceptibility ($\chi^{(2)}$) tensors. 

The common appearances of large in-plane SHG in these films with different OO structures, the absence of CO in PSMO, and the irrelevance of SO in PCMO, rule out most of the symmetry-breaking mechanisms known so far~\cite{CJia,DVEfremov}. The surface-related effects such as piezoelectricity at the (110) surface would produce polarization. However the observed $s$-polarized (in-plane) SHG increases with the film thickness [Fig.~\ref{fig2}(c)], which contradicts to the assumed surface origin. Flexoelectricity induced by the strain gradient in epitaxial films~\cite{DLee} prefers polarization normal to the film. Note that the magnitude of the emerging $\chi^{(2)}$ elements in the PSMO films approach to that of $d_{11}$ of $\alpha$-quartz, which is surprisingly large, leading us to the inquest of the Mn-site polarization (off-centering) with large Born effective charges~\cite{SBhattacharjee,JMRondinelli}.

Deposited on the (110) substrates, the manganite films adopt JT distortions for both $d_{x^{2}-y^{2}}$ and $d_{3x^{2}-r^{2}/3y^{2}-r^{2}}$ OOs, inducing the monoclinic deformation in the lattice~\cite{YWakabayashi} [Fig.~\ref{fig2}(a)]. (Note that the PSMO favors monoclinic 2/$m$ in the bulk OO phase~\cite{HKawano}). When looking into the predicted ferroelectric instability in a cubic perovskite~\cite{JMRondinelli}, the combined unstable phonon modes (second-order JT) are compatible with this shear-mode deformation. In addition, the films are under tensile strain along [1$\bar{1}$0]. Both effects will promote off-centerings of the magnetic $B$-sites, and this atomic displacements are consistent with the lifting of two-fold symmetry along [001] observed in the SHG polarimetry.

\begin{figure}
\includegraphics[scale=0.62]{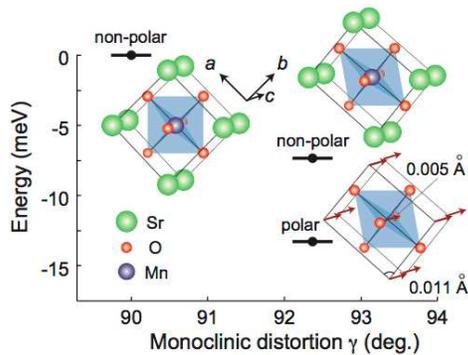}
\caption{\label{fig4} Energy diagram and schematic illustrations for the optimized structures of spin-A-type and half-doped SrMnO$_{3}$ films with and without shear-mode (monoclinic) deformation ($\gamma$). The total energy (per Mn site) is lowered when $\gamma$ becomes larger than 90$^{\circ}$, and further stabilized by the polar displacements of Sr and Mn atoms (indicated by arrows) at $\gamma=$92.37$^{\circ}$ (right bottom, Sr and Mn atoms are abbreviated). The atomic displacements are referred to the center of the O$_{6}$ octahedron, which are mainly along ($b$-$a$) ($\parallel$[$\bar{1}$10]) direction with a slight tilt in the $c$-plane.}
\end{figure}

To confirm this scenario, we examined the stability of the polar structure in the PSMO film by assuming hypothetical spin-A-type and half-doped SrMnO$_{3}$ (Fig.~\ref{fig4}). The structural optimization was performed by first-principles density-functional calculation in the generalized gradient approximation (GGA) with the on-site Hubbard $U$ parameter (GGA+$U$), where a plane-wave basis set with a cutoff of 35 Ry and the ultrasoft pseudopotential were adopted~\cite{calc}. Experimentally determined lattice constants on LSAT(110) with monoclinic crystal distortion ($\gamma=92.37^{\circ}$) in the OO phase were assumed, and 4$\times$4$\times$4 $k$ points were sampled. In addition, an orthorhombic structure ($\gamma=90^{\circ}$) with the same cell volume was checked for comparison. The Hubbard $U$ was chosen to be 4.5 eV for manganese atoms. To find the local potential minima, the polar structures are optimized with the initial displacement of the Mn atoms of 0.2729$\rm{\AA}$ along [1$\bar{1}$0]. As a result, the Mn and Sr atoms found potential minima displaced mainly along [$\bar{1}$10] and slightly to [100] from the center of the O$_{6}$ octahedron for the monoclinic structures (Fig.~\ref{fig4}). These polar displacements are consistent with the experimentally observed SHG; mainly in-plane ($\parallel$[1$\bar{1}$0]) and slightly out-of-plane polarized. The displacement amplitude is the function of the angle of monoclinic distortion, and further enhanced by increasing the lattice constants (not shown) in accord with the former theoretical predictions~\cite{JMRondinelli,SBhattacharjee}. We note that the shift of $A$-site cations, larger in magnitude than that of the $B$-sites, has also been observed for the bulk SBMO in the AFM phase~\cite{HSakai}. The octahedral rotations in the film, as functions of $A$-site element and substrate lattice parameters, will also show strong impacts on these displacive motions, which will be a subject of future study. 

Now the above introduced manganite films are possibly multiferroic, in which the $B$-site off-centering will induce direct couplings of electric, magnetic, and orbital properties to the external fields. We found that the OO transition (and coupled polar states) can be controlled with magnetic field (not shown, cf. Ref.~\cite{YOgimoto}) through the stabilization of the FMM phases. However, the electric-field-control of magnetism has not been successful yet, due mainly to the ``bad insulator'' character of these OO phases. 

It is worth noting that the magnetic information can also be deduced from SHG measurements. The (45$^{\circ},s$) polarized SHG [for the PSMO film, Fig.~\ref{fig1}(a)] partially reflects the magnetization of the film. A kink structure around 260 K, which cannot be explained by the change in the linear optical spectra, indicates the change in magnetic point groups or the slight modification of the orbital and lattice through the magnetostriction. In addition, the SO transition in PCMO film is discernible in the upper panel of Fig.~\ref{fig3}(a), manifesting that the polar state is robust, in contrast to the case of bulk SBMO in which the polarization is largely suppressed by the G-type SO~\cite{HSakai}. To further analyze these magnetic signals, spectroscopic SHG will be helpful. 

To summarize, we realized the engineered polar-magnetic states, possibly produced with the off-centering of the magnetic $B$-site ions, in pseudocubic perovskite manganites by using high-index substrates. The centrosymmetric orbital-orders, $d_{x^{2}-y^{2}}$ and $d_{3x^{2}-r^{2}/3y^{2}-r^{2}}$, drive the shear-mode distortion in the bulk part of the film through the first-order JT distortion constrained on substrates, whereas the $B$-site off-centering is collaborated with the second-order modes. The resultant lattice structure adopts polar point group $m$, which is confirmed by the extensive SHG polarimetry and first-principles calculations. As demonstrated with three different compounds, this would be a general mechanism to produce polarization in magnetic perovskites, and will be useful for applications exploiting efficient cross-correlations.

\begin{acknowledgments}
The authors are grateful to Y. Tomioka and Y. Tokura for providing reference bulk crystals. This work was supported in part by JSPS KAKENHI (21740243) and GCOE for Phys. Sci. Frontier, MEXT, Japan.
\end{acknowledgments}

\newpage 

\begin{thebibliography}{99}

\bibitem{TKimura} T. Kimura {\it et al.,} Nature \textbf{426}, 55 (2003).

\bibitem{YTokura1} Y. Tokura and N. Nagaosa, Science \textbf{288}, 462 (2000).

\bibitem{SWCheong} S. W. Cheong and M. Mostovoy, Nature Mater. \textbf{6}, 13 (2007).

\bibitem{YKonishi} Y. Konishi {\it et al.,}  J. Phys. Soc. Jpn. \textbf{68}, 3790 (1999). 

\bibitem{JHe} J. He, A. Borisevich, S. V. Kalinin, S. J. Pennycook, and S. T. Pantelides, Phys. Rev. Lett. \textbf{105}, 227203 (2010).

\bibitem{NAHill} N. A. Hill, J. Phys. Chem. B \textbf{104}, 6694 (2000).

\bibitem{CJia} C. Jia, S. Onoda, N. Nagaosa, and J. H. Han, Phys. Rev. B \textbf{76}, 144424 (2007). 

\bibitem{DVEfremov} D. V. Efremov, J. van den Brink, and D. I. Khomskii, Nature Mater. \textbf{3}, 853 (2004). 

\bibitem{YTokunaga} Y. Tokunaga {\it et al.,} Nature Mater. \textbf{5}, 937 (2006).

\bibitem{NIkeda} N. Ikeda {\it et al.,} Nature \textbf{436}, 1136 (2005).

\bibitem{EBousquet} E. Bousquet {\it et al.,} Nature \textbf{452}, 732 (2008).

\bibitem{VGopalan} V. Gopalan and D. B. Litvin, Nature Mater. \textbf{10}, 376 (2011).

\bibitem{NABenedek} N. A. Benedek and C. J. Fennie, Phys. Rev. Lett. \textbf{106}, 107204 (2011).

\bibitem{JMRondinelli} J. M. Rondinelli, A. S. Eidelson, and N. A. Spaldin, Phys. Rev. B \textbf{79}, 205119 (2009).

\bibitem{SBhattacharjee} S. Bhattacharjee, E. Bousquet, and P. Ghosez, Phys. Rev. Lett. \textbf{102}, 117602 (2009). 

\bibitem{EBousquet2} E. Bousquet and N. Spaldin, Phys. Rev. Lett. \textbf{107}, 197603 (2011).

\bibitem{HSakai} H. Sakai {\it et al.,} Phys. Rev. Lett. \textbf{107}, 137601 (2011). 

\bibitem{YWakabayashi} Y. Wakabayashi {\it et al.,} J. Phys. Soc. Jpn. \textbf{77}, 014712 (2008).

\bibitem{YOgimoto} Y. Ogimoto {\it et al.,} Phys. Rev. B \textbf{71}, 060403(R) (2005). 

\bibitem{YWakabayashiL} Y. Wakabayashi {\it et al.,} J. Phys: Conf. Ser. \textbf{211}, 012004 (2010).

\bibitem{BBVAken} B. B. van Aken {\it et al.,} Nature \textbf{449}, 702 (2007).

\bibitem{YTomioka} Y. Tomioka, A. Asamitsu, Y. Moritomo, H. Kuwahara, and Y. Tokura, Phys. Rev. Lett. \textbf{74}, 5108 (1995).

\bibitem{MFiebig} M. Fiebig {\it et al.,} Phys. Rev. Lett. \textbf{86}, 6002 (2001).

\bibitem{NOgawa} N. Ogawa {\it et al.,} Phys. Rev. B \textbf{80}, 081106(R) (2009).

\bibitem{KLai} K. Lai {\it et al.,} Science \textbf{329}, 190 (2010).

\bibitem{RRBIRSS} R. R. Birss, {\it Symmetry and Magnetism} (North-Holland Publishing Co., Amsterdam, 1964).

\bibitem{YUozu} Y. Uozu {\it et al.,} Phys. Rev. Lett. \textbf{97}, 037202 (2006). 

\bibitem{DLee} D. Lee {\it et al.,} Phys. Rev. Lett. \textbf{107}, 057602 (2011).

\bibitem{HKawano} H. Kawano {\it et al.,} Phys. Rev. Lett. \textbf{78}, 4253 (1997). 

\bibitem{calc} Here we used the QUANTUM ESPRESSO package developed by S. Baroni {\it et al.}: http://www.pwscf.org/, with the exchange correlation functional introduced by J. P. Perdew, K. Burke, and M. Ernzerhof, Phys. Rev. Lett. \textbf{77}, 3865 (1996).  

\end{thebibliography}

\end{document}